\documentclass[aps,prl,twocolumn,superscriptaddress,nofootinbib]{revtex4-1}

\usepackage{amsmath,amssymb,amsfonts,bm,slashed,graphicx,longtable,color}
\usepackage{multirow, hyperref}
\usepackage{breakurl}

\newcommand{\Jpsi}{J\!/\!\psi}
\newcommand{\Am}[2]{A(#1 \! \to \! \Jpsi #2)}
\newcommand{\AmC}[2]{A_c^{(0)} (#1 \! \to \! \Jpsi #2)}
\newcommand{\AmCn}[2]{A_c^{(1)} (#1 \! \to \! \Jpsi #2)}
\newcommand{\AmU}[2]{A_u^{(0)}(#1 \! \to \! \Jpsi #2)}
\newcommand{\Au}[2]{
\if\relax\detokenize{#2}\relax
      A^{(#1)}_{u}
  \else
    A^{(#1)}_{u, #2 }
  \fi
}
\newcommand{\Ac}[2]{
\if\relax\detokenize{#2}\relax
      A^{(#1)}_{c}
  \else
    A^{(#1)}_{c, #2 }
  \fi
}
\newcommand{\ac}{\mathcal{A}_c}
\newcommand{\Gam}[2]{\bar{\Gamma}(#1 \! \to \! \Jpsi #2)}
\newcommand{\Br}[2]{\ensuremath{\mathcal{B}(#1 \! \to \! \Jpsi #2)}}
\newcommand{\bls}{\bar{\lambda}^2}
\newcommand{\Ru}{R_u}
\newcommand{\rb}{\bar{\rho}}
\newcommand{\eb}{\bar{\eta}}

\newcommand{\im}{\mbox{Im}}
\newcommand{\re}{\mbox{Re}}
\newcommand{\Scp}{S_{\!f}}
\newcommand{\Ccp}{C_{\!f}}
\newcommand{\phs}{{\phantom{*}}}

\newcommand{\Bbar}{\,\overline{\!B}{}}
\newcommand{\Dbar}{\,\overline{\!D}{}}
\newcommand{\Kbar}{\,\overline{\!K}{}}
\def\B0bar{\Bbar{}^0}
\def\D0bar{\Dbar{}^0}
\def\K0bar{\Kbar{}^0}
\def\ov{\overline}
\newcommand{\nn}{{\nonumber}}

\allowdisplaybreaks

\begin{document}

\title{\boldmath Towards more precise determinations of the CKM phase \texorpdfstring{$\beta$}{beta} }

\author{Zoltan Ligeti}
\affiliation{Ernest Orlando Lawrence Berkeley National Laboratory,
University of California, Berkeley, CA 94720}

\author{Dean J.\ Robinson}
\affiliation{Ernest Orlando Lawrence Berkeley National Laboratory,
University of California, Berkeley, CA 94720}
\affiliation{Department of Physics, University of California, Berkeley, CA 94720}

\begin{abstract}
We derive a new flavor symmetry relation for the determination of the weak phase
$\beta=\phi_1$ from time-dependent CP asymmetries and $B \to J/\psi P$ decay rates. In
this relation, the contributions to $\sin2\beta$ proportional to $V_{ub}$ are
parametrically suppressed compared to the contributions in the $B \to J/\psi K^0$ time-dependent CP asymmetry alone. This
relation uses only $SU(3)$ flavor symmetry, and does not require further
diagrammatic assumptions. The current data either fluctuate at the $2\sigma$
level from expectations, or may hint at effects of unexpected magnitude from contributions proportional to $V_{ub}$ or from isospin breaking.
\end{abstract}

\maketitle

\section{Introduction}

CP violation in $B \to \Jpsi K_S$ will be measured at the percent level at
Belle~II~\cite{Belle2predictions} and LHCb~\cite{LHCb-PUB-2014-040}, a precision
several times better than today~\cite{Aubert:2009aw, Adachi:2012et,
Aaij:2015vza}, and crucial for improving the sensitivity to new physics in $B$
mixing (see, e.g.,~\cite{Charles:2013aka}). This projected uncertainty is
comparable to the characteristic size of the CKM-suppressed uncertainties,
proportional to $V_{ub}$,\footnote{In the literature this is often referred to
as ``penguin pollution". Since we are not concerned with diagrammatic arguments,
we identify terms by CKM factors.} in the time-dependent CP asymmetry,
\begin{align}
	\label{eqn:DCS}
	&\frac{ \Gamma[\Bbar(t) \!\to\! f] \! - \! \Gamma[B(t) \!\to\! f]}
 		 {\Gamma[\Bbar(t) \!\to\! f] \! + \! \Gamma[B(t) \!\to\! f] }
 		 = \Scp \sin(\Delta m\, t) - \Ccp \cos(\Delta m\, t) , \nn\\
	&\Scp = \frac{2\,\mbox{Im}[(q/p)(\ov A_f/A_f)]}{1 + |\ov A_f/A_f|^2}\,,\quad
 		 \Ccp = \frac{1 - |\ov A_f/A_f|^2}{1 + |\ov A_f/A_f|^2}\,.
\end{align}
Here $f$ denotes final states composed of $J/\psi$ and a pseudoscalar meson,
$P$; $A_f = \langle f | {\cal H} |B^0\rangle$, $\ov A_f = \langle f | {\cal H}
|\B0bar\rangle$; $\Delta m$ is the mass difference between the two neutral $B$
mass eigenstates, $|B_{H,L}\rangle = p | B^0 \rangle \mp q |\B0bar\rangle$; and 
we neglect the small ${\cal O}(\Delta \Gamma/\Gamma,\, |q/p| - 1)$ effects in
the $B_d$ system, as well as ${\cal O}(\epsilon_K)$ effects, which are
straightforward to include~\cite{Grossman:2002bu}.

At the current level of precision, the relation
\begin{equation}\label{eqn:SPP}
	S_{K_S} =  \sin(2\beta) 
	+ \mathcal{O}\big[ V_{ub}^* V^\phs_{us}/ (V_{cb}^*V^\phs_{cs} )\big]
	+ \ldots  \,,
\end{equation}
truncated at leading order has been sufficient to extract the CKM phase $\beta
\equiv \arg[-V^*_{cb}V^\phs_{cd} / (V_{tb}^*V^\phs_{td})]$. The theoretical
uncertainty is limited by our ability to compute or bound the subleading
contribution to the decay amplitude, proportional to $V_{ub}$. This is the
$A_u$ term in the decay amplitude,
\begin{equation}
	\label{eqn:ACU}
	A = \lambda_c^{q}\,A_c + \lambda_u^{q}\,A_u\,, \qquad 
	\lambda_{i}^q \equiv V^*_{ib}V^{\phantom{*}}_{iq}\,,
\end{equation}
($i = u,c$ and $q = d,s$), which has a different weak phase and possibly a
different strong phase than the dominant $A_c$ term.

The upcoming experimental precision has renewed interest in constraining the
effects of this ``\emph{$V_{ub}$ contamination}"  in measurements of $\beta$ and
its analog in $B_s$ decays, $\beta_s$. Comparisons between $B_d \to \Jpsi
\rho^0$ and $B_s \to \Jpsi \phi$~\cite{Fleischer:1999zi, Aaij:2014vda} rely both
on flavor symmetry and diagrammatic arguments.  It has also been proposed to use
$B_s \to \Jpsi K_S$ to control the $V_{ub}$ term in $B_d \to \Jpsi K_S$  (see,
e.g., Ref~\cite{DeBruyn:2014oga}). Other approaches attempt to constrain the
$V_{ub}$ contribution from global fits to multiple observables using flavor
$SU(3)$~\cite{Ciuchini:2005mg,Faller:2008zc,Ciuchini:2011kd, Jung:2012mp,
Jung:2012vd}, often with additional simplifying assumptions, or attempt to
compute the corresponding hadronic matrix element using QCD factorization (see,
e.g., Ref~\cite{Frings:2015eva}).  Some of these works claim that the $V_{ub}$
contamination can be enhanced to several percent, which is challenged by a lower
estimate of rescattering effects using measured rates~\cite{Gronau:2008cc}.

In this paper we derive a flavor $SU(3)$ relation for $\beta$, involving the
$B_d \to \Jpsi K_S$, $B_d \to \Jpsi \pi^0$, $B^+ \to \Jpsi K^+$ and $B^+ \to
\Jpsi \pi^+$ branching ratios and CP asymmetries, in which, in the $SU(3)$
limit, the contributions linear in $V_{ub}$ cancel. This permits extraction of
$\beta$ up to parametrically suppressed contributions, compared to the $V_{ub}$
contamination in Eq.~\eqref{eqn:SPP}.  Our results rely only on group theoretic
relations among the decay amplitudes, and do not involve diagrammatic or
factorization arguments. The same relations imply a lower bound for the
presently unmeasured $B_s \to \Jpsi \pi^0$ decay rate.

\section{Amplitude Relations}

We obtain $SU(3)$ relations for the $B \to \Jpsi f$ decay amplitudes by
application of a Wigner-Eckart expansion, after embedding the Hamiltonian and
the in- and out-states into $SU(3)$ representations. The $B$ in-states furnish a
flavor anti-triplet, $[B_3]_i = (B^+, B_d, B_s)$. The charmless pseudoscalar
out-states furnish a singlet, $[P_1] = \eta_1$, and the usual octet,
\begin{equation}
	\label{eqn:BVJ}
	[P_8]^i_j = \begin{pmatrix} \frac{\pi^0}{\sqrt{2} } + \frac{\eta_8}{\sqrt{6} } & \pi^+ & \phantom{-}K^{+} \\[5pt]
 	\pi^- & - \frac{\pi^0}{\sqrt2} + \frac{\eta_8}{\sqrt{6} }  & \phantom{-}K^{0} \\[5pt]
	  K^{-} & \bar{K}^{0} & - \frac{2\eta_8}{\sqrt{6} } \end{pmatrix}.
\end{equation}
We allow an arbitrary $\eta$-$\eta'$ mixing angle, such that the mass
eigenstates are $\eta^{(\prime)} = \eta_8\cos\theta \mp \eta_1 \sin \theta$.

The effective Hamiltonian for $B \to J/\psi P$ decay contains four-quark
operators that mediate $\bar b \to \bar q^i q_j \bar q^k$ or $\bar b \to c \bar
c \bar q^i$ transitions ($q = u,d,s$). Under $SU(3)$ flavor, this embeds into
$\bm{3} \otimes \bar{\bm{3}} \otimes \bm{3} = \bm{3} \oplus \bm{3}' \oplus
\bar{\bm{6}} \oplus \bm{15}$ irreducible representations.  The nonzero
independent components of the Hamiltonian are given in Eq.~(53) of Ref.~\cite{Grossman:2013lya}.
Finally, $SU(3)$ and isospin breaking is encoded by insertions of the usual
octet spurions, $[\mathcal{M}]^i_j \equiv \ \varepsilon\,\mbox{diag}\{1,1,-2\}$ and
$\delta\,\mbox{diag}\{1,-1,0\}$ respectively.

We work to first order in $G_F$ and to all orders in $\alpha_s$. In the $SU(3)$
limit, the $A_c$ and $A_u$ terms in Eq.~(\ref{eqn:ACU}) each depend on three
reduced matrix elements, corresponding to the $\bm{3}$, $\bar{\bm{6}}$, and
$\bm{15}$ pieces of the Hamiltonian. For $A_c$, the $\bar{\bm{6}}$ and $\bm{15}$ terms only arise
from electroweak penguin contributions, suppressed by $\alpha_{\textrm{em}}$. 
These are accounted together with other sources of isospin breaking in
$A_c$, which are comparable in size.  The electroweak penguin contributions to
$A_c$ transforming as the $\bm{3}$ (which probably dominate) are automatically
absorbed in the leading $A_c$ contributions.

The decay amplitudes are expanded to $\mathcal{O}(\varepsilon^p)$ via
\begin{align}
	\label{eqn:WET}
& \Am{B}{f} = \sum_{w,p} X^{p}_w (C^p_w)_{B;f}\,,\\
& (C^p_w)_{B;f} \equiv \frac{\partial^2}{\partial f \partial B}
	\bigg[[P_{1,8}]^{i_1\ldots}_{j_1\ldots} \mathcal{H}^{p_1\ldots}_{q_1\ldots} 
	\big([\mathcal{M}]^{k_1}_{l_1}\cdots\big) [B_3]_r\bigg]_w.
\nn
\end{align}
Here $w$ labels a set of linearly independent $SU(3)$ tensor contractions,
$\mathcal{H}$ is the Hamiltonian, and there are $p$ insertions of $\mathcal{M}$.
The $X^{p}_w$ are reduced matrix elements, while $C^{p}_w$ encode the weak
physics, $p$th order $SU(3)$ breaking effects, and group theoretic factors. 
Finding $SU(3)$ sum rules at order $\varepsilon^p$ is equivalent to computing
kernels of $(C^p_w)_{B;f}$~\cite{Grossman:2012ry, Grossman:2013lya}.

It is useful to derive relations that hold independently for the  $A_c$ and
$A_u$ amplitudes in Eq.~\eqref{eqn:ACU}. In anticipation of the need to account
for $SU(3)$ breaking effects, we further expand each reduced matrix element
order-by-order in $SU(3)$ breaking, and write
\begin{equation}
	\label{eqn:AE}
	A_c = \Ac{0}{} + \varepsilon \Ac{1}{} + \ldots\,, \quad A_u = \Au{0}{} + \varepsilon \Au{1}{} + \ldots\,.
\end{equation}
In the $SU(3)$ limit, we have
\begin{subequations}\label{eqn:7all}
\begin{align}
	0 & = \AmC{B_s}{\pi^0} \label{eqn:ACSRPI1}\,, \\ 
	\ac 
	& \equiv \AmC{B_d}{K^0}  = \AmC{B^+}{K^+} \nn \\
	& = \AmC{B^+}{\pi^+} = \AmC{B_s}{\bar{K}^0} \nn \\
	& = -\sqrt{2}\AmC{B_d}{\pi^0}\,,\label{eqn:AC}
\end{align}
\end{subequations}
Hereafter, we write $\ac$ instead of the $\Ac{0}{}$ amplitudes in
Eq.~\eqref{eqn:AC}. Considering the first order $SU(3)$ breaking contributions
to the amplitudes independently, we find
\begin{subequations}
\label{eqn:ACSRNP}
\begin{align}
	0 & = \AmCn{B_s}{\pi^0}\,, \label{eqn:ACSRNPI1}\\
	0 & = \sqrt{2}\AmCn{B_d}{\pi^0} + \AmCn{B^+}{\pi^+}\,,  \label{eqn:ACSRNPI2}\\
	0 & = \AmCn{B_d}{K^0} - \AmCn{B^+}{K^+}\,,  \label{eqn:ACSRNPI3} \\
	0 & = \AmCn{B^+}{K^+} + \AmCn{B^+}{\pi^+} \nn \\*
	& \qquad + \AmCn{B_s}{\bar{K}^0}\,, \label{eqn:ACSRNPI4}
\end{align}	
\end{subequations}
Equations \eqref{eqn:ACSRNPI1}--\eqref{eqn:ACSRNPI3} are isospin
relations, and hold to all orders in the $SU(3)$ breaking 
parameter~$\varepsilon$. Finally, the $A_u$ amplitudes in the $SU(3)$ limit satisfy~\cite{Grossman:2003qp,
Savage:1989ub, Zeppenfeld:1980ex}
\begin{subequations}
\label{eqn:AUSRP}
\begin{align}
	0 & = \AmU{B^+}{\pi^+} - \AmU{B^+}{K^+}\,,\label{eqn:AUSRP1}\\
	0 & = \AmU{B_d}{K^0} - \AmU{B_s}{\bar{K}^0}\,, \label{eqn:AUSRKK} \\
	0 & = \sqrt{2}\AmU{B_d}{\pi^0} - \sqrt{2}\AmU{B_s}{\pi^0} \nn \\
	& \qquad + \AmU{B_d}{K^0}\,. \label{eqn:AUSRP3}
\end{align}	
\end{subequations}
Besides Eqs.~\eqref{eqn:7all}--\eqref{eqn:AUSRP}, there are further relations
involving $\Jpsi \eta^{(\prime)}$ states, that are not needed for our analysis.
Similar relations also hold for vector mesons, with obvious replacements.

It is often assumed based on diagrammatic arguments that the $\AmU{B_s}{\pi^0}$
contribution in Eq.~\eqref{eqn:AUSRP3} can be neglected (see, e.g.,
\cite{Ciuchini:2005mg,Faller:2008zc,Ciuchini:2011kd, Jung:2012mp}). We make no
such assumption.  The current limits on $\AmU{B_s}{\pi^0}$ are weak, in the
sense that the data allows this contribution to be sizable. Below we use
Eq.~\eqref{eqn:AUSRP3} to set a lower bound on the branching ratio
$\Br{B_s}{\pi^0}$.

\section{\boldmath relation for \texorpdfstring{$\sin(2\beta)$}{s2b}}

Given the flavor symmetry relations, we proceed to construct an $SU(3)$ relation
among branching ratios and time-dependent CP asymmetries, that permits
extraction of $\beta$ without $V_{ub}$ contamination in the $SU(3)$ limit. This
relation will only involve $B_d$ or $B^+$ decays, so hereafter we denote $A_f
\equiv A(B \to \Jpsi f)$, for $B = B_d$, $B^+$.

Besides the $SU(3)$ and isospin breaking parameters,
\begin{equation}
	\varepsilon \sim \frac{f_K}{f_\pi}-1\, \sim 0.2,\qquad
	\delta \sim \frac{m_d-m_u}{\Lambda_{\chi\textrm{SB}}} \lesssim 1\%\,,
\end{equation}
we also expand certain observables in
\begin{equation}
\bls \equiv -\frac{\lambda_u^s}{\lambda_c^s}
  \frac{\lambda_c^d}{\lambda_u^d} \simeq 0.05\,, \quad
  \Ru \equiv \sqrt{\rb^2 + \eb^2} \simeq 0.37 \,,
\end{equation}
where $\rb + i\eb \equiv -\lambda_u^d/\lambda_c^d \simeq 0.15 + 0.34\,i$ is the
apex of the unitarity triangle. Powers of $\Ru$ track powers of $V_{ub}$, and
enter with corresponding powers of $A_u/A_c$. We make no assumptions concerning
the size of $|A_u/A_c|$. While $\varepsilon$ and $\Ru$ are not particularly
small parameters, $\Ru^2$, $\varepsilon\Ru$ and $\varepsilon^2$ can be treated
as $\ll 1$. We therefore expand physical observables to this order, and seek
relations without $\mathcal{O}(\varepsilon, \Ru)$ terms.

Expanding to next-to-leading order in these small parameters, the CP-averaged
rate is
\begin{align}\label{eqn:ARPE} 
	& \Gam{B}{f} = \big[|\vec{p}_{B \to \Jpsi f}| /(8\pi m_B^2)\big]\,
	\big|\lambda_c^q\big|^2\, \big|\Ac{0}{f}\big|^2\nn\\
	&\quad \times \bigg[1 + 2\varepsilon\, \re \frac{\Ac{1}{f}}{\Ac{0}{f}} 
	+ 2\re\frac{\lambda_u^q}{\lambda_c^q}\, \re \frac{\Au{0}{f}}{\Ac{0}{f}}
	+ \ldots \bigg] .
\end{align}
Corrections are $\mathcal{O}(\Ru^2, \varepsilon\Ru, \varepsilon^2)$ and
$\mathcal{O}(\varepsilon\Ru\bls, \varepsilon^2)$ in $b\to d,s$ processes
respectively. The $\varepsilon A_c^{(1)}/A_c^{(0)}$ terms arise from first order
$SU(3)$ breaking and must be kept, as they are parametrically larger than ${\cal
O}(\bls)$. Note they do not satisfy the same relations as  the $A_c^{(0)}$
terms.

Applying Eqs.~\eqref{eqn:AC} and \eqref{eqn:ACSRNPI3} to Eq.~\eqref{eqn:ARPE} yields
\begin{align}\label{eqn:DKK}
\Delta_{K}  & 
	\equiv \frac{\Gam{B_d}{K^0} - \Gam{B^+}{K^+} }{\Gam{B_d}{K^0}  + \Gam{B^+}{K^+} } \\
& = \re \frac{\lambda_u^s}{\lambda_c^s}\,\re \frac{\sqrt{2}\Au{0}{K_S} - \Au{0}{K^+}}{\ac}
	 + \mathcal{O}(\varepsilon \Ru \bls,\, \delta)\,. \nn
\end{align}
We emphasize that the $\varepsilon^n\, \re\big[\Ac{n}{} / \Ac{0}{}\big]$ terms
in Eq.~\eqref{eqn:ARPE} are canceled up to isospin breaking corrections. We
have also made the replacement $\Am{B_d}{K^0} = \sqrt{2}\Am{B_d}{K_S}$.
Analogously, we also obtain
\begin{align}\label{eqn:DPP}
\Delta_{\pi}  
	& \equiv \frac{2\Gam{B_d}{\pi^0} - \Gam{B^+}{\pi^+}}{2\Gam{B_d}{\pi^0} + \Gam{B^+}{\pi^+}} \\
& = -\re \frac{\lambda_u^d}{\lambda_c^d} \, \re \frac{\sqrt{2}\Au{0}{\pi^0} + \Au{0}{K^+}}{\ac}
	+ \mathcal{O}(\Ru^2, \varepsilon\Ru,\, \delta)\,, \nn
\end{align}
where we replaced $\Au{0}{\pi^+}\!$ with $\Au{0}{K^+}$ using
Eq.~\eqref{eqn:AUSRP1}.

The CP asymmetry in $B_d \to \Jpsi f$ can be written as
\begin{equation}\label{eqn:SCPE}
\Scp =
  - \eta_{f} \bigg[ \sin 2\beta + 2\,\im\frac{\lambda_u^q}{\lambda_c^q}\,
  \re \frac{\Au{0}{f}}{\Ac{0}{f}}\, \cos 2\beta\, + \ldots\bigg] ,
\end{equation}
where $\textrm{CP}|J/\psi\, f\rangle = \eta_f|J/\psi\, f\rangle$, and
corrections are $\mathcal{O}(\Ru^2, \varepsilon \Ru)$ and
$\mathcal{O}(\varepsilon\Ru \bls)$ for $b \to d,s$ respectively.  The $\re
\big[\Au{0}{} / \Ac{0}{}\big]$ term in Eq.~\eqref{eqn:SCPE} dominates the
$V_{ub}$ contamination in Eq.~\eqref{eqn:SPP}. From Eq.~\eqref{eqn:SCPE} the CP
asymmetries for $B_d \to \Jpsi K_S$ and $B_d \to \Jpsi \pi^0$ are
\begin{equation}
\label{eqn:SCPKP}
\begin{split}
	S_{K_S} - \sin 2\beta
	 & = 2\, \im \frac{\lambda_u^s}{\lambda_c^s}\, \re \frac{\sqrt{2}\Au{0}{K_S}}{\ac} \cos 2\beta + \ldots , \\
	S_{\pi^0} +\sin2\beta 
	 & = 2\,\im \frac{\lambda_u^d}{\lambda_c^d}\, \re \frac{\sqrt{2}\Au{0}{\pi^0}}{\ac} \cos2\beta+\ldots .
\end{split}
\end{equation}  

Eliminating the $V_{ub}$ contamination -- the $\Au{0}{}$ terms -- in Eqs.~(\ref{eqn:DKK}), (\ref{eqn:DPP}), and (\ref{eqn:SCPKP}), one obtains the relation
\begin{align}
\label{eqn:MRP} 
& (1 + \bls) \sin 2\beta 
	= S_{K_S} - \bls S_{\pi^0}\\
& \quad - 2\big(\Delta_{K} + \bls\Delta_{\pi}\big) \cos 2\beta\, \tan\gamma 
+ \mathcal{O}(\varepsilon\Ru \bls, \Ru^2 \bls, \delta)\,, \nn
\end{align}
where $\gamma \equiv \arg(-\lambda_u^d/\lambda_c^d)$.  Eq.~\eqref{eqn:MRP} is
the main result of this paper. In the $SU(3)$ limit, the $V_{ub}$ contamination
in $S_{K_S}$, $\Delta S_{K_S} \equiv S_{K_S} - \sin 2\beta$, is canceled by
contributions from $\Delta_K$, $\Delta_\pi$ and $S_{\pi^0}$. This leaves only
corrections parametrically higher order in $\varepsilon$, $\delta$ or $\Ru$, 
\begin{equation}
	\varepsilon\Ru \bls\, \re \frac{\Au{0}{}}{\ac}\,, \quad
	\Ru^2\, \bls\,\bigg|\frac{\Au{0}{\pi}}{\ac}\bigg|^2, \quad 
	\delta\,\re \frac{\Ac{\delta}{K}}{\ac} \,,
\end{equation}
where $\delta \Ac{\delta}{K}$ is the isospin breaking difference of $A_{c,K^0}$
and $A_{c,K^+}$, arising in $\Delta_K$.  

The $\mathcal{O}(\varepsilon\Ru \bls)$ $SU(3)$-breaking correction in
Eq.~\eqref{eqn:MRP} is unambiguously smaller than the $V_{ub}$ contamination in
$\Delta S_{K_S}$, of order ${\cal O} ( \Ru \bls)$.

The $\mathcal{O}(\Ru^2 \bls)$ terms in Eq.~\eqref{eqn:MRP} are dominated by the
$V_{ub}^2$ terms in $\Delta_\pi$, which are numerically enhanced by $\tan \gamma
\simeq 2.6$. If $A_u/A_c = {\cal O}(1)$, then these corrections are not
numerically suppressed, since $\Ru\tan\gamma \simeq 0.9$.  However, in this
case, future data should show an enhancement of $\Delta_\pi$ compared to its
present value (see Table \ref{tab:PV}), which will constrain this possibility. 
If $A_u/A_c \ll 1$ then this $\mathcal{O}(\Ru^2 \bls)$ correction is negligible.

Concerning the isospin breaking ${\cal O}(\delta)$ contribution to
Eq.~\eqref{eqn:MRP}, if $A_u/A_c = {\cal O}(1)$ and $\delta\, \re[\Ac{\delta}{}
/ \ac] \sim 1\%$, then this term is subleading compared to $\Delta S_{K_S}$. If
$A_u/A_c \ll 1$ and $\delta\, \re[\Ac{\delta}{} / \ac] \sim 1\%$, then this term
may be numerically larger than $\Delta S_{K_S}$. However, in this case, the
experimental upper bound on $\Delta_K$ should decrease.  It may also be
possible to obtain constraints on the isospin violating matrix element
$\Ac{\delta}{K} / \ac$ using other methods, in order to extract $\beta$ from
Eq.~\eqref{eqn:MRP} at sub-percent precision.

\section{Numerical results and predictions}

The four observables in Eqs.~(\ref{eqn:DKK}), (\ref{eqn:DPP}), and
(\ref{eqn:SCPKP}) depend on $\beta$ and the real parts of the three
$\Au{0}{f}/\ac$ amplitude ratios.  We may therefore extract these matrix
elements and $\beta$ from a fit to these four observables, noting one may also
extract $\beta$ directly from Eq.~\eqref{eqn:MRP}. We use the SM fit values $\gamma
= 67^\circ \pm 2^\circ$ and $\bls \simeq 5.36\times
10^{-2}$~\cite{Hocker:2001xe, *Charles:2004jd} as inputs, and determine $\Ru$
from the identity $\Ru \equiv \sin \beta/\sin(\gamma + \beta)$.  The SM CKM
fit results for $\Ru$ (or $\rb$ and $\eb$) are not used, as they depend strongly
on the assumption of negligible $V_{ub}$ contamination in $\beta$, whereas the
SM fit result for $\gamma$ has only a small dependence on the direct $\beta$
measurement.

\begin{table}[tb]
\tabcolsep 15pt
\begin{tabular}{ll}
\hline
Observable  &  \multicolumn{1}{c}{Measurement} \\
\hline\hline
\Br{B_d}{K^0}  &  $(8.63 \pm 0.35)\times 10^{-4}$ \\
\Br{B^+}{K^+}  &  $(10.28 \pm 0.40)\times 10^{-4}$ \\
$\Delta_{K}$  &  $-(5.0 \pm 2.8)\times 10^{-2}$\\
\hline
\Br{B_d}{\pi^0}  &  $(1.74 \pm 0.15)\times 10^{-5}$  \\
\Br{B^+}{\pi^+}  &  $(4.04 \pm 0.17)\times 10^{-5}$ \\
$\Delta_{\pi}$  &  $-(3.7 \pm 4.8)\times 10^{-2}$ \\
\hline
$S_{K_S}$  &  $0.682 \pm 0.021$  \\
$S_{\pi^0}$  &  $-0.93 \pm 0.15$  \\
\hline
\end{tabular}
\caption{The experimental measurements used \cite{Amhis:2014hma}.}
\label{tab:PV}
\end{table}

The experimental data for these observables are shown in Table~\ref{tab:PV} from
HFAG~\cite{Amhis:2014hma}.  The $S_{K_S}$ value is the average of $S_{J/\psi
K_S}$ from BaBar, Belle, and LHCb, with other charmonium states $\psi(2S)$,
$\chi_c$, etc., excluded, since those hadronic matrix elements are not related
by $SU(3)$.  One then finds from Eq.~(\ref{eqn:MRP})
\begin{equation}	
	\label{eqn:NBV}
	\beta = 27.8^\circ \pm 2.9^\circ\,, 
\end{equation}
and from Eqs.~(\ref{eqn:DKK}), (\ref{eqn:DPP}), and (\ref{eqn:SCPKP}), the
matrix elements $\re[\Au{0}{K^+}/\ac] = -0.4 \pm 0.4$,  $\re[\sqrt{2}
\Au{0}{\pi^0}/\ac] = 0.2 \pm 0.3$, $\re[\sqrt{2}\Au{0}{K_S}/\ac] = -5.5 \pm
2.3$, and $\Ru/\Ru^{\textrm{SM}} = 1.3 \pm 0.1$.

The $\pi^0$ matrix element is consistent with recent global fits or QCD
factorization analyses (see, e.g., Refs.~\cite{DeBruyn:2014oga,
Frings:2015eva}). On the other hand, $\re[\sqrt{2}\Au{0}{K_S}/\ac]$ is
larger than the expected size of $V_{ub}$ contamination or isospin breaking.
This arises from the large central value of the linear combination
\begin{equation}\label{delta_exp}
\Delta_{K} + \bls \Delta_{\pi} = -0.052 \pm 0.028\,.
\end{equation}
Assuming that the $V_{ub}$ contamination and isospin violation are small, so
that $\beta$ takes its current SM fit value, $\beta = (21.9 \pm
0.8)^\circ$~\cite{Charles:2004jd}, then Eq.~\eqref{eqn:MRP} and the  $S_{K_S}$
and $S_{\pi^0}$ data predict,
\begin{equation}\label{delta_pred}
	\Delta_{K} + \bls \Delta_{\pi} = 0.001 \pm 0.009\,.
\end{equation}
The source of the $2\sigma$ tension between Eqs.~\eqref{delta_exp} and
\eqref{delta_pred} is the same  as that between Eq.~\eqref{eqn:NBV} and the SM
fit for $\beta$.  Future higher statistics data for the CP averaged $B \to \Jpsi
K$ and $B \to \Jpsi \pi$ rates, together with the time dependent CP asymmetries
in $B_d \to \Jpsi K^0$ and $B_d \to \Jpsi \pi^0$, is required to resolve this
tension.\footnote{Future measurements of these rates may require
combined analyses with other decays, to simultaneously constrain the
isospin asymmetries and the $B^+B^-$ versus $B_d\bar{B}_d$ production in
$\Upsilon(4S)$ decay.  Current analyses either assume isospin symmetry to
measure the production rate difference, or assume equal production rates to
measure the branching ratios entering $\Delta_{K,\pi}$~\cite{Amhis:2014hma, Agashe:2014kda}.}

Combining Eq.~\eqref{eqn:AUSRP3} with Eqs.~\eqref{eqn:DKK} and \eqref{eqn:DPP},
one finds in the $SU(3)$ limit,
\begin{equation}
	\label{eqn:BSDR}
	\Delta_K + \bls\,\Delta_\pi = \re \frac{\lambda_u^s}{\lambda_c^s}\,\re \frac{\sqrt{2} \AmU{B_s}{\pi^0}}{\ac}\,.
\end{equation}
The sizable experimental central value for the left-hand side [cf.\
Eq.~\eqref{delta_exp}] is therefore connected to the possibility of a sizable
amplitude $\AmU{B_s}{\pi^0}$.  According to  Eq.~\eqref{eqn:ACSRPI1},
$\AmC{B_s}{\pi^0}$ vanishes by isospin.  Neglecting the possibility of
cancellations between $\AmU{B_s}{\pi^0}$ and the isospin violating contribution
to $A_c(B_s \to \Jpsi\pi^0)$, Eq.~\eqref{eqn:BSDR} implies the lower bound
\begin{equation}\label{eqn:bound}
\frac{\Gam{B_s}{\pi^0}}{\Gam{B}{K}} \ge 
  \frac{(\Delta_{K} + \bar\lambda^2 \Delta_{\pi})^2}{2\cos^2\gamma} \,,
\end{equation}
where we neglected small phase space differences. From the current
experimental data in Table~\ref{tab:PV}, we obtain
\begin{equation}
	\label{eqn:BSLB}
	\Br{B_s}{\pi^0} \ge 4.4\times 10^{-6}\,,
\end{equation}
at the $1\sigma$ level, and $> 1.1\times10^{-6}$ at the $90\%$ CL. This is to be compared to the SM expectation of $\mathcal{O}(10^{-7})$. The
experimental uncertainties dominate this result, and are larger than the
theoretical uncertainty in Eq.~(\ref{eqn:bound}).

\begin{figure}[t]
\includegraphics[width = .9\linewidth]{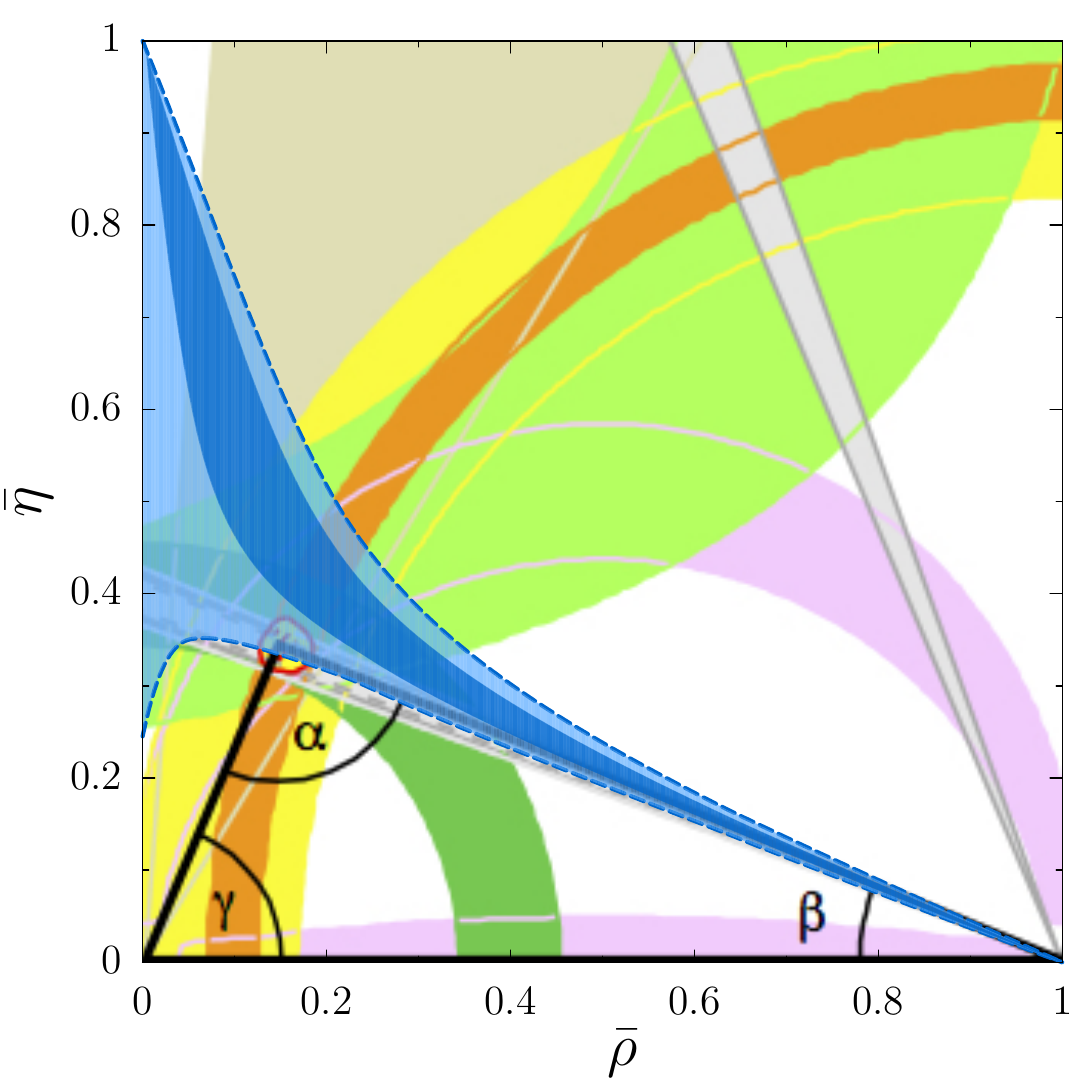}
\caption{Constraint from Eq.~\eqref{eqn:MRP} at $\pm 1 \sigma$ (transparent dark
blue) and $\pm 2\sigma$ (transparent light blue) in the $(\bar{\rho},
\bar{\eta})$ plane, overlaid on the SM CKM fit~\cite{Charles:2004jd}.}
\label{fig:UT}
\end{figure}

One can use Eq.~\eqref{eqn:MRP} to derive an allowed region in the
$(\bar{\rho},\bar{\eta})$ plane.  In Fig.~\ref{fig:UT} we show this constraint
from the current data, compared to other bounds. The sizable uncertainty of
$\Delta_{K}$ leads to a somewhat loose constraint.  The $\pm 1 \sigma$ range at
present tends to favor a slightly larger $\beta$, and is in better agreement
with measurement of $|V_{ub}|$ from inclusive rather than exclusive semileptonic
$B$ decays. More precise measurements of $S_{K_S}$, $S_{\pi^0}$, $\Delta_{K}$,
and $\Delta_{\pi}$ are needed to improve the statistical significance of this
constraint and to decide if there is an interesting tension with the SM CKM
fit. 

Future data will also give other means to explore whether the uncertainties in
$\beta$ are under control and to gain confidence about bounds on the $V_{ub}$
contamination.  For example: (i)~The $\Delta_\pi$ observable in
Eq.~(\ref{eqn:DPP}) only receives an $\Au{0}{}$ contribution from the $\bm{15}$
representation, so more precise data can be used to constrain the size of this
matrix element, which also contributes to $\Delta S_{K_S}$; (ii)~The direct CP
asymmetries can be used to extract the imaginary parts of the $\Au{0}{f}/\ac$
amplitude ratios, which provide a lower bound on the $|A_u/A_c|^2$ terms in
$\Delta_\pi$; (iii)~When $|V_{ub}|$ measurements improve, comparison of the SM
CKM fit excluding $S_{K_S}$ with Eq.~\eqref{eqn:MRP} will provide independent
information on possible origins of the tension in Fig.~\ref{fig:UT}.

\section{Acknowledgments}

This work was supported in part by the Office of High Energy Physics of the
U.S.\ Department of Energy under contract DE-AC02-05CH11231 (ZL) and
by the NSF under grant No. PHY-1002399 (DR).


%

\end{document}